\RequirePackage{fix-cm}
\documentclass[twocolumn]{svjour3}          
\smartqed

\usepackage{graphicx}
\usepackage{xcolor}
\usepackage{lipsum}
\usepackage{amsmath}
\usepackage{bm}
\usepackage{hyperref}
\usepackage{ar}
\usepackage{cite}

\newcommand{\ub}{{\bf u}}

\newcommand{\Xb}{{\bf X}}

\newcommand{\Fb}{{\bf F}}

\begin{document}

\title{Flowing fibers as a proxy of turbulence statistics}

\author{Marco E. Rosti         \and
        Stefano Olivieri   \and
        Arash A. Banaei \and
        Luca Brandt        \and 
        Andrea Mazzino
}

\institute{M.E. Rosti \and A.A. Banaei \and L. Brandt \at Linn\'{e} Flow Centre and SeRC (Swedish e-Science Research Centre), KTH Mechanics, SE 100 44 Stockholm, Sweden \\ \email{merosti@mech.kth.se}
 \and
S. Olivieri \and A. Mazzino \at DICCA, University of Genova, Via Montallegro 1, 16145 Genova, Italy
\and
S. Olivieri \and A. Mazzino \at INFN, Genova Section, Via Dodecaneso 33, 16146 Genova, Italy\\ \email{andrea.mazzino@unige.it} 
}

\date{Received: date / Accepted: date}

\maketitle

\begin{abstract}
The flapping states of a flexible fiber fully coupled to a three-dimensional turbulent flow are investigated via state-of-the-art numerical methods. Two distinct flapping regimes are predicted by the phenomenological theory recently proposed by Rosti et al. [Phys. Rev. Lett. 121, 044501, 2018]: the under-damped regime, where the elasticity strongly affects the fiber dynamics, and the over-damped regime, where the elastic effects are strongly inhibited. In both cases we can identify a critical value of the bending rigidity of the fiber by a resonance condition, which further provides a distinction between different flapping behaviors, especially in the under-damped case. We validate the theory by means of direct numerical simulations and find that, both for the over-damped regime and for the under-damped one, fibers are effectively slaved to the turbulent fluctuations and can therefore be used as a proxy to measure various two-point statistics of turbulence. Finally, we show that this holds true also in the case of a passive fiber, without any feedback force on the fluid.
\end{abstract}

\section{Introduction} \label{intro}

The interaction between elastic structures and fluid \sloppy{flows} concerns several problems relevant for both biological and industrial applications, such as, e.g., flow control~\cite{rosti_kamps_bruecker_omidyeganeh_pinelli_2017a}, passive locomotion of swimmers~\cite{fish2006review,shelley2011flapping,lacis2014passive} and energy harvesting from flow-induced vibrations~\cite{peng2009,boragno2012elastically,li2016review,olivieri2017fluttering,boccalero2017power}. Moreover, fundamental studies have focused on the dynamics of constrained, elastic filaments in laminar flows, investigating the essence of flapping instabilities and related phenomena~\cite{shelley2011flapping,zhang2000flexible,huang_shin_sung_2007a,bagheri2012symmetry}.

On the other hand, in the field of particle-laden flows attention has been devoted to the dynamics of fiber-like objects dispersed in laminar or turbulent flows. In this latter framework, a further distinction can be made depending on the size of the fiber compared to that of the existing flow scales. A significant number of contributions have considered fibers with length smaller than the Kolmogorov scale~\cite{voth2017anisotropic,sabban2017temporally}, revealing the nature of effects such as preferential alignment for neutrally-buoyant rods~\cite{parsa2012rotation} and preferential concentration when their inertia enters into play~\cite{marchioli2010,ardekani2017}.

Fewer studies have addressed the case of finite-size fibers whose length is comparable to scales belonging to the inertial range. Among these, experimental analyses have outlined the potential of using dispersed rigid rods as a measurement tool for turbulent flows~\cite{parsa2014inertial}, since their tumbling rate is found to be approximately equal to the characteristic time of turbulent structures of comparable size (provided inertia does not affect the fiber dynamics~\cite{bordoloi2017rotational,bounoua2018tumbling}).

The case of flexible fibers has been recently addressed  both experimentally~\cite{brouzet2014polymer,verhille2016} and numerically~\cite{kunhappan2017,rosti2018flexible}. One of the main findings of Ref.~\cite{rosti2018flexible} is the identification of a flapping regime where the fiber deformation is slaved to turbulent fluctuations, enabling to quantify their statistical properties by measuring only the distance and velocity difference between the fiber free ends.

Further attempts of such a Lagrangian description have seen the employment of other kinds of particles for evaluating both two-point (limited to distances between particles smaller than the Kolmogorov viscous scale) and single-point quantities~\cite{kramel2016chiral,hejazi2017}, paving the way for new strategies to investigate turbulent flows. Overall, such efforts are needed in order to increase our understanding of turbulence and establish, in particular, a connection between scaling laws and spatial structures, e.g.~vortex filaments \cite{siggia_1981,douady1991}. In passive scalar turbulence, this connection leads to a complete understanding of the meaning of intermittency and anomalous scaling~\cite{vergassola1997,falkovich2001}.

The goal of this work is to present an exhaustive description of the dynamical phenomenology associated to a long flexible fiber (i.e. having a rest length falling within the turbulence inertial range of scales) freely moving in a controlled turbulent flow. More specifically, we aim to categorize the different regimes that can be predicted theoretically by combining a simple structural model for the fiber with a widely studied turbulence model: the so-called homogeneous, isotropic and stationary turbulence model. We will present and discuss the different flapping states that may occur depending on the choice of the characteristic parameters, supporting our physical intuitions with evidence from fully-resolved numerical simulations. Furthermore, the underlying hypothesis of passive fiber, on which the phenomenological model that will be introduced in this work relies, will be reviewed by considering numerical experiments in which the feedback exerted by the fiber on the flow is deactivated; This will clarify whether this is crucial to capture the essential dynamics.

The idea of using a fiber to measure two-point turbulence statistics was recently proposed by Rosti et al.~\cite{rosti2018flexible}. Our goal here is to give a detailed presentation on how to exploit a flexible fiber as a proxy of turbulence statistics and to present new results.

The present work is structured as follows: Sec.~\ref{sec:method} presents the numerical method along with its validation. In Sec.~\ref{sec:model} we introduce our phenomenological model while in Sec.~\ref{sec:flapping-states} we discuss the different dynamical regimes and provide corroborations from DNS. Sec.~\ref{sec:passive} concerns the case of passive fiber and in Sec.~\ref{sec:experiments} we propose parameters for possible experiments. Finally, Sec.~\ref{sec:conclusions} draws some conclusions.

\section{Numerical method}
\label{sec:method}

We consider the fully coupled dynamics of a flexible fiber, governed by the Euler-Bernoulli equation, immersed in an incompressible three-dimensional turbulent flow field, governed by the Navier-Stokes equations.

In an inertial, Cartesian frame of reference the equations of momentum and mass conservation for the incompressible flow read as
\begin{align} \label{eq:NS}
  \frac{\partial u_i}{\partial t} + \frac{\partial u_i u_j}{\partial x_j} &= - \frac{1}{\rho} \frac{\partial p}{\partial x_i} + {\nu} \frac{\partial^2 u_i}{\partial x_j x_j} + f_i^\mathrm{T} + f_i^\mathrm{F}, \\
  \frac{\partial u_i}{\partial x_i} &= 0,
  \label{eq:NS2}
\end{align}
where $u_i$ is the fluid velocity field, $p$ the pressure, $f_i^\mathrm{T}$ and $f_i^\mathrm{F}$ two volume body forces used to sustain the turbulent flow and to account for the presence of the immersed fiber, respectively, and $\rho$ and $\nu$ the density and kinematic viscosity of the fluid (being $\mu=\rho \nu$ the dynamic viscosity). The problem can be made non-dimensional by choosing reference length and velocity scales, $U_*$ and $L_*$, and by defining the Reynolds number as $Re=\rho U_* L_*/\mu$. The equations of motion are solved in a triperiodic box, with periodic boundary conditions applied in all the three Cartesian directions.

The forcing $f_i^\mathrm{T}$ is used to generate and sustain a fully turbulent flow with homogeneous, isotropic, and stationary statistics. To do so, we use the spectral forcing scheme proposed by Eswaran and Pope \cite{eswaran_pope_1988a}, which involves the addition of energy to the Fourier modes of the velocity at wavenumbers inside a low wavenumber shell. The injected energy is obtained from a formulation based on Uhlenbeck-Ornstein processes and the resulting flowfield displays neither anisotropy nor unreasonably high correlation times \cite{eswaran_pope_1988a}.

The fluid-solid coupling force $f_i^\mathrm{F}$ is obtained by the Immersed Boundary Method (IBM), a technique used to simulate the flow past solid bodies, first developed by Peskin \cite{peskin_1972a} to simulate blood flow inside a heart. The main feature of this method is that the numerical grid does not need to conform to the geometry of the object, which is replaced by the body force distribution $f_i^\mathrm{F}$ which mimics the presence of the body on the fluid and restores the desired velocity boundary conditions on the immersed surfaces. Two sets of grid points are needed: a fixed Eulerian grid $x_i$ for the fluid and a moving Lagrangian grid $X_i$ for the structure. The body force is found by first computing the fluid-solid interaction force as
\begin{equation} \label{eq:FSI}
F_i=\frac{U_i^\mathrm{ib}-U_i^\Gamma}{\Delta t},
\end{equation}
where $U_i^\mathrm{ib}$ is the interpolated fluid velocity on the Lagrangian points which does not satisfy the boundary condition on the immersed objects, $U_i^\Gamma$ the desired velocity of the Lagrangian points, and $\Delta t$ the time step. The interpolation from the Eulerian grid to the Lagrangian one of the fluid velocity is performed using a smooth Delta function, i.e.
\begin{equation}
U_i^\mathrm{ib}=\int_V u_i~\delta \left( X_i-x_i \right)~dV,
\end{equation}
where the integration is performed over the whole fluid domain $V$. Similarly, the spreading of the fluid-solid interaction force (Eq. \eqref{eq:FSI}) from the Lagrangian grid to the Eulerian one is performed as
\begin{equation}
f_i^\mathrm{F}=\rho_1 \int_{c} F_i~\delta \left( X_i - x_i \right)~ds,
\label{eq:spread}
\end{equation}
where $s$ is the curvilinear coordinate along the fiber. The update of the Lagrangian points is achieved by solving a separate equation that describes the dynamics of the flexible fiber; in our simulations we use the Euler-Bernoulli beam equation together with the inextensibility condition \cite{segel_2007a}
\begin{align}
\rho_1 \frac{\partial^2 X_i}{\partial t^2} &= \frac{\partial}{\partial s} \left( T \frac{\partial X_i}{\partial s}\right) - \gamma \frac{\partial^4 X_i}{\partial s^4} - F_i,\label{eq:EB}\\
 \frac{\partial X_i}{\partial s} \frac{\partial X_i}{\partial s} &= 0,
\end{align}
where $T$ is the tension, $\rho_1$ the the fiber linear density, $\gamma$ the bending rigidity, and $F_i$ the fluid-solid interaction force. This model is fully justified as far as the ratio between the filament length and its diameter, $\AR = c/d$, is much larger than unity. The fiber is free to move in the flow, thus, zero force, torque and tension boundary conditions are enforced at the two ends of the fiber. Gravitational effects are neglected, i.e.,~the Froude number is always much larger than unity.

The fluid equations are solved numerically on a staggered grid, with pressure points located at the cell center and velocity components at the cell faces, using a second order finite difference code. Eqs. \eqref{eq:NS} and \eqref{eq:NS2} are advanced in time by a fully explicit fractional step-method, where the third-order Runge-Kutta method is used, and the Poisson equation is solved by Fast Fourier Transform. To solve the fiber equation we follow the explicit two-step method proposed by Huang et al.~\cite{huang_shin_sung_2007a}.

Note that, the exact relation between the fiber volume and linear density is not clearly defined in the method described above due to the uncertain definition of the shape and cross-section of the fiber, which, although being mono-dimensional in the theory, has a finite thickness due to the spreading operation (Eq.~\eqref{eq:spread}), with the Dirac-delta function having a support of 4 grid points. The problem can be solved as follows: we first simulate a free fiber with a prescribed bending rigidity $\gamma$ in void, i.e.~without fluid, and measure its main oscillation frequency $f_\mathrm{osc}$. By standard normal mode analysis techniques, we can write that $f_\mathrm{osc} = \alpha \sqrt{\gamma/(\rho_1 c^4)}$ (where $\alpha \approx \pi/22.4$), which can then be used to obtain the actual value of the fiber linear density due to its finite thickness. 

\subsection{Validation}
\begin{figure*}
  \centering
  \includegraphics{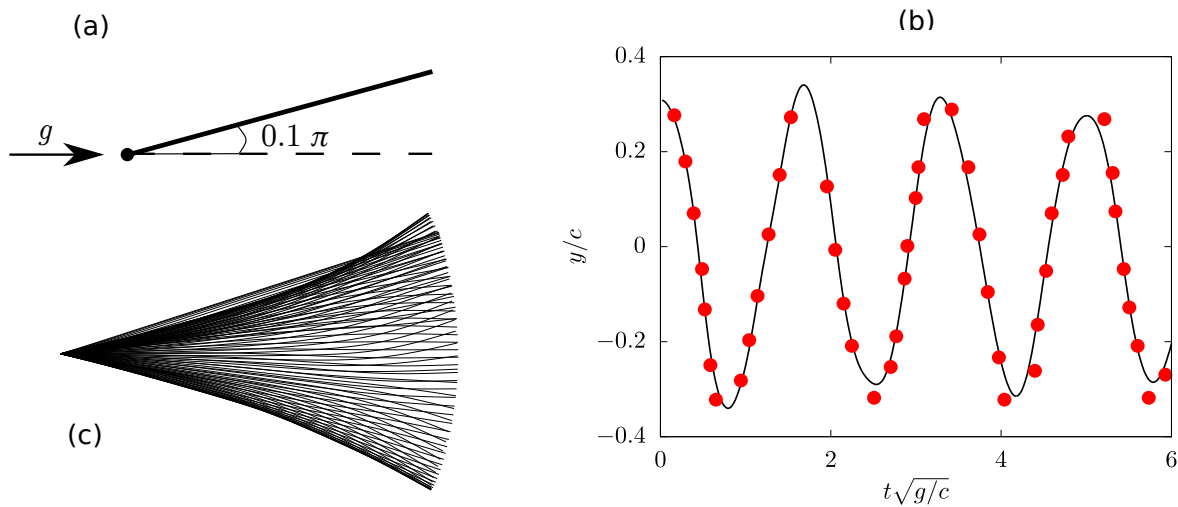}
  \caption{Validation of the hanging filament under gravity: (a) sketch of the initial condition; (b) time evolution of the $y$-coordinate of the free-end of a filament due to gravity (the solid black line represents our numerical results, while the red dots those by Huang et al.~\cite{huang_shin_sung_2007a}); (c) envelope of filament positions over time.}
\label{fig:val1}
\end{figure*}

\begin{figure*}
  \centering
  \includegraphics{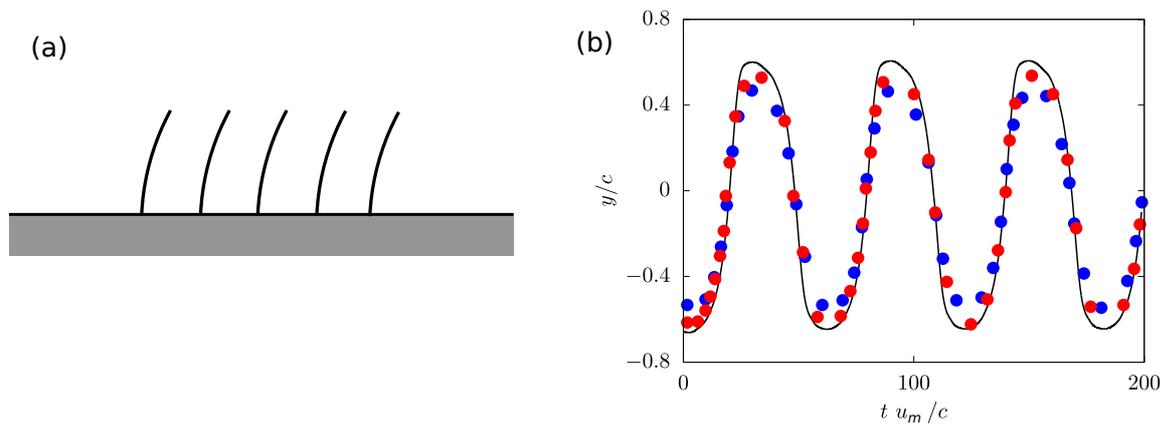}
  \caption{Validation of filaments in pulsating channel flow: (a) sketch of the case; (b) the streamwise displacement of the tip of the last filament at the right end of the row with respect to time. The solid black line is used to show our numerical results, while the red and blue dots experimental and numerical results from the literature \cite{pinelli_omidyeganeh_bruecker_revell_sarkar_alinovi_2016a}.}
\label{fig:val2}
\end{figure*}

The numerical code used in this work has been extensively validated in the past for turbulent flow simulations \cite{rosti_brandt_2017a,rosti_izbassarov_tammisola_hormozi_brandt_2018a,shahmardi_zade_ardekani_poole_lundell_rosti_brandt_2019a}. Here, we provide one more comparison with literature results for the specific case of a flexible filament.

First, we validate the structural solver by studying the oscillations of a hanging filament under gravity, as done by Huang et al.~\cite{huang_shin_sung_2007a}. The filament is initially held stationary with an angle of $0.1\pi$ from the vertical (Fig.~\ref{fig:val1}a) and then, after being released, starts swinging due to the gravity force. The unit filament is discretised in our simulation with $100$ Lagrangian points. Fig.~\ref{fig:val1}b shows the time history of the free-end position obtained by our simulation (solid line) and by Huang et al.~\cite{huang_shin_sung_2007a} (red dots); a good agreement with the literature data is evident.

Next, we validate the fluid-structure interaction solver by considering a pulsating flow in a plane channel filled with glycerine ($Re=u_\textrm{m} c/\nu = 60$ being $u_\textrm{m}$ the maximum velocity). A row of $5$ filaments is hinged vertically on the bottom wall (Fig.~\ref{fig:val2}a). The pulsating frequency of the channel is $0.016 c/u_\textrm{m}$, matching the filaments natural frequency (the filaments Young modulus is $2.05/\rho u_\textrm{m}^2$). Fig.~\ref{fig:val2}b shows the streamwise displacement of the tip of the last filament with respect to time: our results (solid line) are compared with the experimental measurements (red dots) and with the simulations (blue dots) reported by Pinelli et al.~\cite{pinelli_omidyeganeh_bruecker_revell_sarkar_alinovi_2016a}. Both the frequency of oscillation and the magnitude of the displacement match the literature results.

\subsection{DNS setup}
\begin{figure}
  \includegraphics[width=0.49\textwidth]{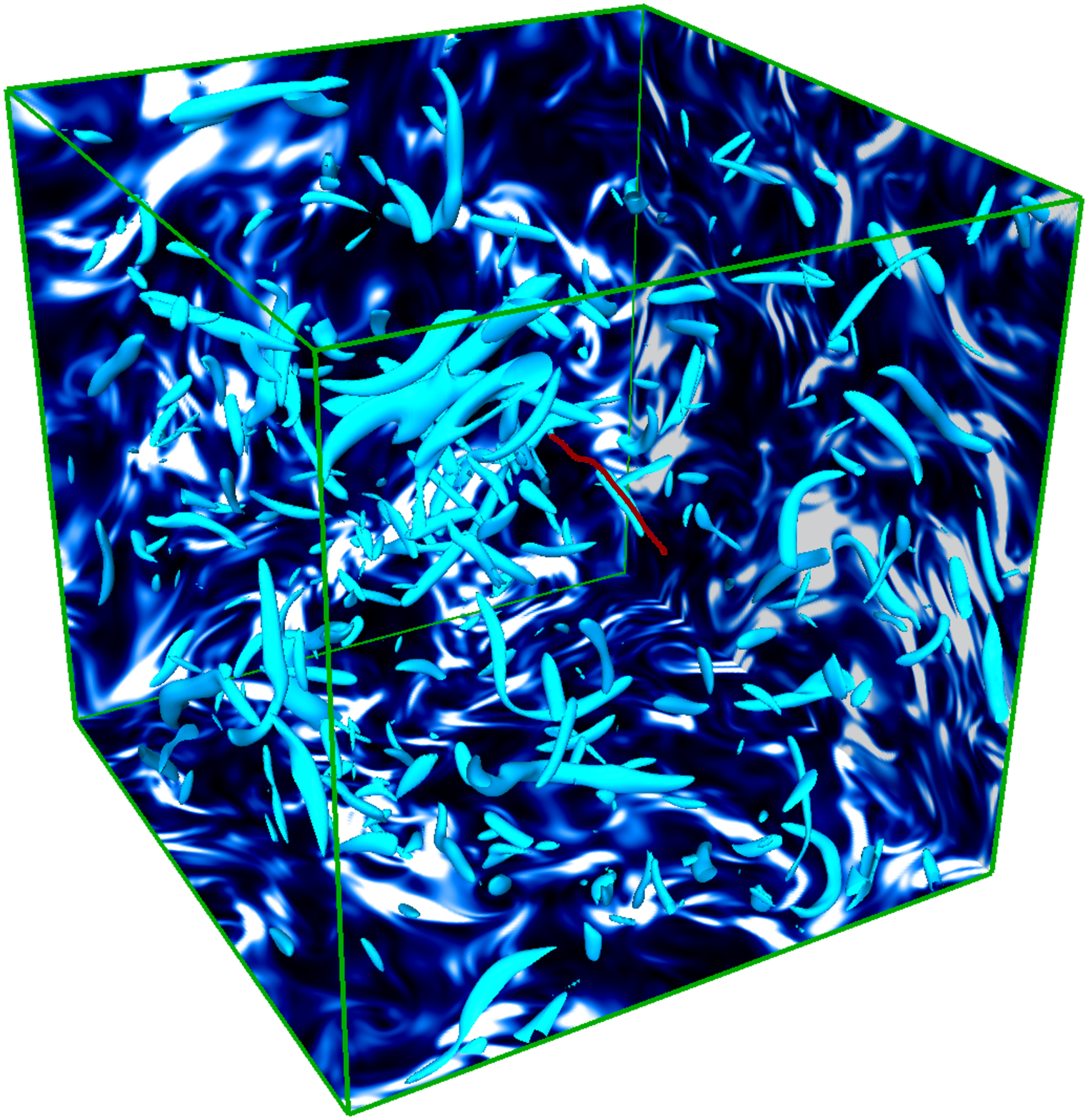}
  \caption{A snapshot from our DNS of homogeneous isotropic turbulence in which a flexible fiber (line in dark red) is immersed. Isosurfaces of $\mathcal{Q}$ (in cyan) depict the istantaneous vorticity field, and the three back planes are coloured according to the value of the enstrophy field.}
\label{fig:etc}
\end{figure}
Details are given here on the numerical simulations whose data are reported in the remaining part of the paper. We consider a flexible fiber immersed in an incompressible three-dimensional turbulent flow field; in Fig.~\ref{fig:etc} we report an instantaneous visualization of the turbulent flow along with the dispersed fiber to give a qualitative insight of the resulting scenario.

The equations of motion are solved in a triperiodic box with size $L=2\pi$, discretized on a Cartesian uniform mesh using $128$ grid points per side. With this grid resolution, the resulting turbulence two-point statistics is consistent with the $\frac{4}{5}$th Kolmogorov law in the inertial range of scales (as one can notice for the Eulerian quantities in Fig.~\ref{fig:S23p}), with negligible differences if doubling the number of nodes in all directions. The resulting turbulent dissipation rate $\epsilon$, made dimensionless with the cube of the velocity root-mean square divided by the size of the box, is about $2.6$ and the Reynolds number at the Taylor microscale is about $Re_\lambda=92$. Concerning the discretization of the fiber, the spacing between the Lagrangian points is taken equal to that of the Eulerian grid described before. Depending on the chosen length of the fiber, the number of Lagrangian points used in the simulations ranges from $16$ to $41$.

The results presented hereafter are obtained by solving the equation of motion and tracking the fibers dynamics for more than $40$ large-eddy turnover times, with the statistical quantities averaged over at least $20$ large-eddy turnover times.

\begin{figure}
  \includegraphics[width=0.49\textwidth]{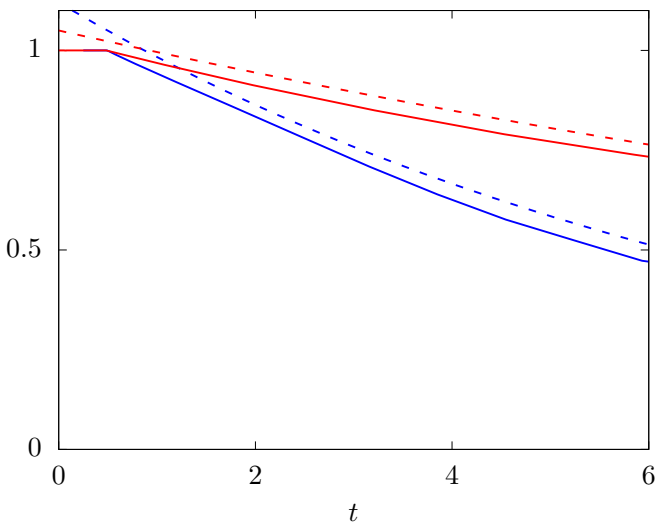}
  \caption{Time history of the normalized velocity (red) and angular velocity (blue) of a filament translating and rotating in a quiescent fluid. The dashed lines are analytical expressions in the form of $c_1 e^{t/t_0}$ plotted with the values of $t_0$ that best fit our data and shifted upwards for major clarity.}
\label{fig:stokes}
\end{figure}

\begin{figure*}
 \centering
  \includegraphics{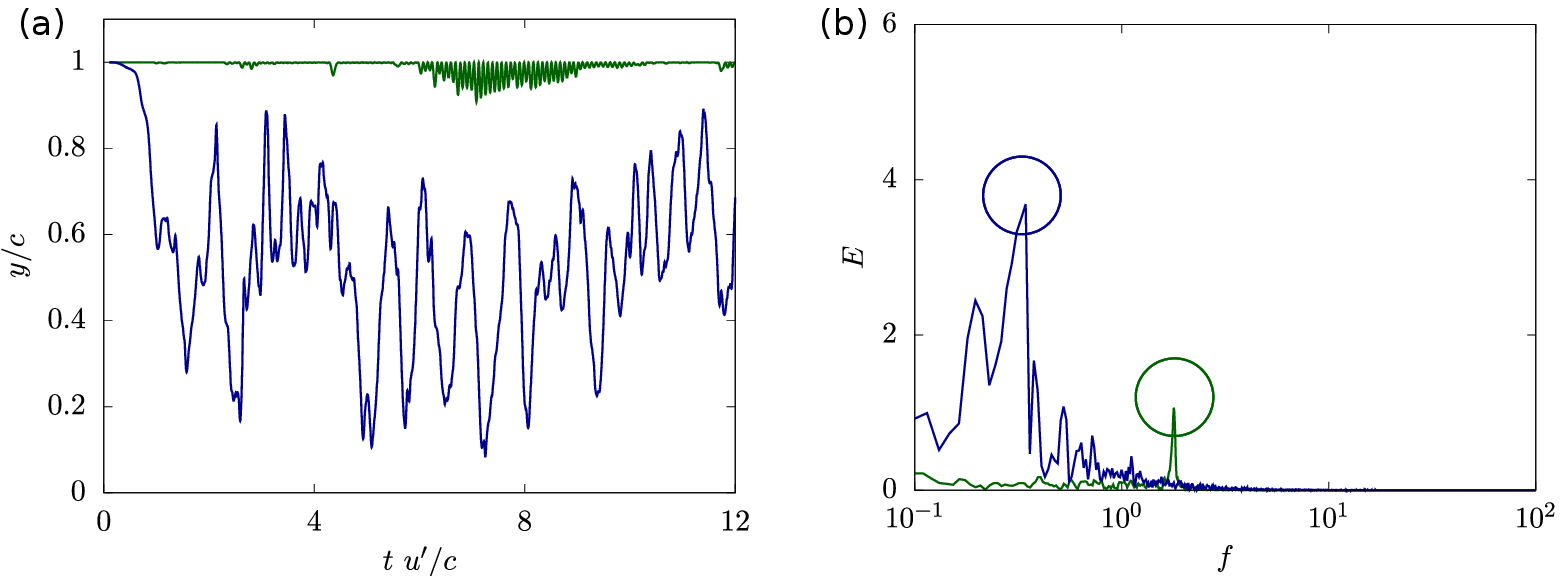}
  \caption{(a) Time histories of the end-to-end distance and (b) corresponding spectra for the subcritical case $\gamma / \gamma^{\mbox{\tiny{ud}}}_{\mbox{\tiny{crit}}} = 0.3$ (blue) and the supercritical case $\gamma / \gamma^{\mbox{\tiny{ud}}}_{\mbox{\tiny{crit}}} = 50$ (green) in the under-damped regime. In (b), the dominant frequencies are marked with a circle and, for the sake of visualisation, values of the supercritical case are divided by $10^3$. }
  \label{fig:end2end}
\end{figure*}

\section{Phenomenological model} \label{sec:model}
In this section, we start by reviewing the arguments presented in Ref.~\cite{rosti2018flexible}. Our goal is to model the coupling between the fiber and the flow in a simple, yet effective way. For this purpose, we begin by estimating the different timescales of the problem.

Recalling the fiber dynamical equation~\eqref{eq:EB} and neglecting at first the forcing from the flow, one can define the fiber elastic time $\tau_{\gamma}$ by balancing the inertial and bending terms:
\begin{equation}
\tau_{\gamma}=\alpha \left( \frac{\rho_1c^4}{\gamma} \right)^{1/2},
\label{eq:tauGamma}
\end{equation} 
where $c$ is the fiber length, $\rho_1$ the linear density difference, $\gamma$ the bending rigidity and the factor $\alpha \approx \pi/22.4$ results from a normal mode analysis, corresponding to the first natural frequency of the beam in the unsupported (free-free) case.
If we model now the fluid-solid interaction force $\Fb$ with a simple viscous term $ \Fb=-\mu (\dot \Xb - \ub)$~\cite{Cox70}, another characteristic time can be estimated by balancing the inertial term with the introduced damping:
\begin{equation}
\tau_{\mu}=\frac{2\rho_1}{\mu}.
\end{equation} 
It has to be noted that the expression chosen for the drag does not account for anisotropic effects (as done, e.g., for fibers in low-Reynolds flows~\cite{YS07}), since in our case the flow has an essentially isotropic nature and no preferential alignment is expected.

An analogy can be drawn between our model and a damped harmonic oscillator, so that we can derive an expression for the equivalent damping ratio:
\begin{equation}
\zeta=\frac{\tau_{\gamma}}{\tau_{\mu}}=\frac{\alpha c^2\mu}{2\rho_1^{1/2}\gamma^{1/2}}. 
\end{equation}
The critical condition $\zeta=1$ represents the threshold between two different regimes: for $\zeta<1$ (\textit{under-damped} regime) the elasticity is expected to strongly affect the fiber dynamics, while for $\zeta>1$ (\textit{over-damped} regime) elastic effects are strongly inhibited. More details will be given in the next section.

Concerning the flow, from the well-known $\frac{4}{5}$th Kolmogorov law combined with dimensional analysis, the turnover time of turbulent eddies of size $r$ is expected to behave as
\begin{equation}
\tau(r) \sim r^{2/3}{\epsilon}^{-1/3},
\end{equation} 
where $\epsilon$ is the turbulent dissipation rate of kinetic energy. Let us now assume that the flow structures (i.e. eddies) effectively acting in the fluid-structure coupling are those with the same size of the fiber, i.e. $r \approx c$. This also implies that the activated oscillation mode of the fiber is essentially the first one (whose associated timescale is supplied by Eq.~\eqref{eq:tauGamma}). Having collected these relations, we are able to make predictions regarding the flapping states of the elastic structure depending on the physical parameters.

\begin{figure*}
\centering
 \includegraphics{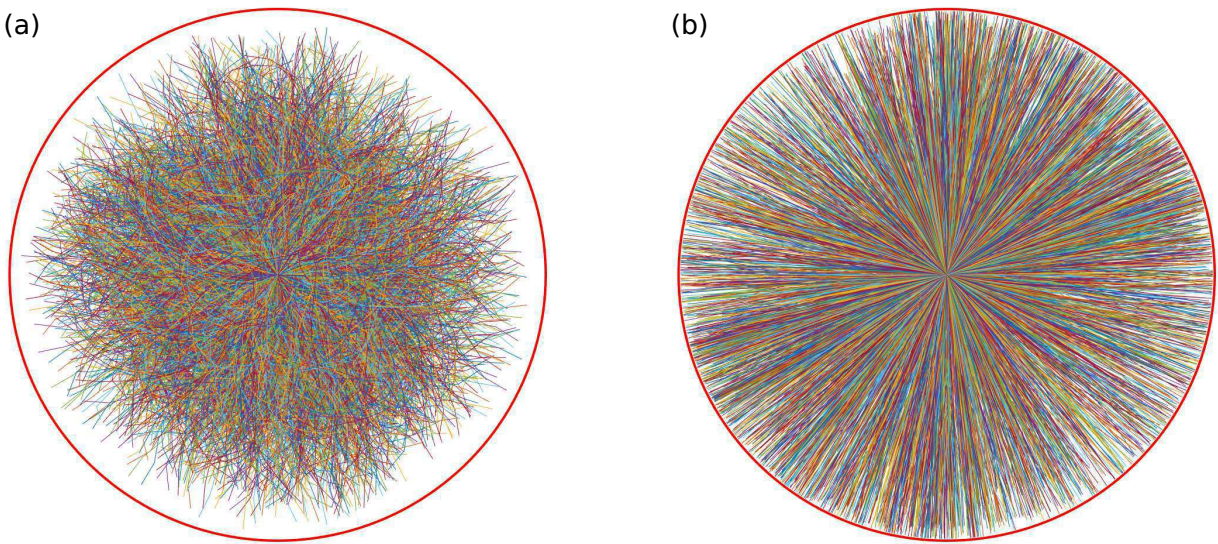}
  \caption{Superposition of instantaneous fiber configurations for the subcritical case $\gamma / \gamma^{\mbox{\tiny{ud}}}_{\mbox{\tiny{crit}}} = 0.3$ (left) and the supercritical case $\gamma / \gamma^{\mbox{\tiny{ud}}}_{\mbox{\tiny{crit}}} = 50$ (right) in the under-damped regime (same as Fig.~\ref{fig:end2end}). The red circle encloses the undeformed fiber configuration.}
\label{fig:def}
\end{figure*}
\section{Fiber flapping states} \label{sec:flapping-states}
\subsection{Under-damped regime} \label{sec:ud}
For $0<\zeta<1$ (under-damped regime), we expect that the fiber response shows an oscillatory behavior. Therefore, we impose a resonance condition between the elastic time and the hydrodynamic one, i.e. $\tau_{\gamma} \approx \tau(c)$, from which a critical value of the bending rigidity can be found:
\begin{equation}
\gamma^{\mbox{\tiny{ud}}}_{\mbox{\tiny{crit}}}\sim c^{8/3}\epsilon^{2/3} \rho_1 \alpha^2. 
\label{gamma-crit-ud}
\end{equation}
Looking at the expression above, a further distinction can be made. In the limit of vanishing $\gamma$ (\textit{subcritical} case), the fiber can be thought to be slaved to the flow due to the relatively faster forcing compared with its response, therefore flapping at the eddy frequency. Such a slaved dynamics is expected to hold true when both the rotational and translational Stokes number are sufficiently small. The rotational and translational Stokes numbers are defined as
\begin{equation}
\mathit{St} = \frac{t_0 u_0}{c_0},
\label{eq:stokes_rot}
\end{equation}
where $t_0$ is the Stokes time and $c_0$ and $u_0$ are characteristic length and velocity scales. Two different length and velocity scales are used: for the translational case the length scale is set equal to the fiber length $c_0=c$ and the velocity scale is chosen equal to the rms fluid velocity $u_0 = u'$, while for the rotational Stokes number the length scale is set equal to $c_0=\pi c$  and the velocity scale to half the velocity difference $u_0=\delta u_{\parallel}/2$ at the fiber scale. The two Stokes times are measured through numerical simulations where we consider a rotating and translating fiber with speed $u_0$ in a quiescent fluid with the same viscosity and density used in the rest of the work. We then measure the decay time from our results, as shown in Fig.~\ref{fig:stokes}. The values of both these two non-dimensional parameters, computed using the expressions given above, are approximately $\mathcal{O}(1)$. In particular, the translational Stokes number for $c/L=0.16$ and $\rho_1/\left( \rho_0 L^2 \right) = 0.042$ is equal to $7$ while the rotational one equals $0.3$. The fact that the translational Stokes number is not smaller than unity, as it is for the rotational Stokes number, suggests that the possibility of using a fiber as a proxy of two-point statistics of turbulence is controlled by the (smallness of the) rotational Stokes number rather than by the translational Stokes number. Further analysis of this issue are however worth investigating to arrive at a firm conclusion.

In the opposite \textit{supercritical} case, where the elastic time is much smaller than the hydrodynamic one, the fiber reaction is expected to be far more rapid than the fluid forcing.

To corroborate these expectations, we consider results from DNS of two cases, both belonging to this under-damped regime but with different $\gamma$. We start by looking at how the fiber end-to-end distance varies in time (Fig.~\ref{fig:end2end}a). The two curves look clearly different: in the subcritical case, finite-size and continuous variations of the signal are found, with a mean end-to-end value of around $0.55$; conversely, in the supercritical case, the fiber remains almost unbent (the mean end-to-end distance is about $0.99$) and isolated bursts are observed (such as that occurring from $t u'/c \approx 6$, where $u'$ is the velocity root-mean square), when interactions occur with energetic eddies. These observations are confirmed by the pictorial views in Fig.~\ref{fig:def} where the overlapped fiber configurations at different time instants are collected: we observe appreciable deformations in the subcritical case, while the fiber essentially behaves like a rigid rod in the supercritical regime.

For a deeper understanding, the corresponding temporal spectra are analysed and reported in Fig.~\ref{fig:end2end}b. A substantial difference can be noticed between the two cases: in the supercritical case, the dominant frequency is the natural frequency of the fiber, i.e. $ f = 1/\tau_\gamma$, with a clear peak in the spectra, while for the subcritical one the peak of the spectrum is less distinct and found at the turbulent frequency $f_{\mbox{\tiny{turb}}} = 1/\tau(c)$.

\begin{figure}
  \centering
  \includegraphics[width=0.49\textwidth]{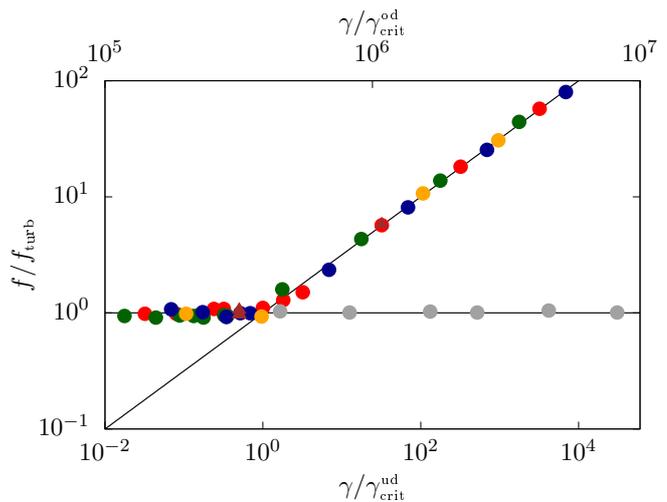}
  \caption{The fiber oscillation frequencies (normalized by the turbulence frequency at the fiber length scale) as a function of the fiber bending rigidity (normalized by the corresponding critical value, given by Eq. \eqref{gamma-crit-ud} for the under-damped case (lower axis) and Eq. \eqref{gamma-crit-od} for the over-damped case (upper axis)). 
Blue: $c/L=0.12$ and $\rho_1/(\rho_0 L^2)=0.042$;
red: $c/L=0.16$ and $\rho_1/(\rho_0 L^2)=0.042$;
green: $c/L=0.20$ and $\rho_1/(\rho_0 L^2)=0.042$; 
orange: $c/L=0.16$ and $\rho_1/(\rho_0 L^2)=0.014$ or 0.125; 
gray: $c/L=0.16$ and $\rho_1/(\rho_0 L^2)=0$;
brown triangles:  $c/L=0.16$ and $\rho_1/(\rho_0 L^2)=0.042$, passive cases.}
  \label{fig:freq}
\end{figure}

To confirm the theoretical predictions, we explore the parametric space by considering fibers with different combinations of density $\rho_1$, length $c$ and bending rigidity $\gamma$. Fig.~\ref{fig:freq} shows the ratio between the dominant frequency $f$ and the turbulent frequency $f_{\mbox{\tiny{turb}}}$ as a function of the ratio between the fiber bending stiffness $\gamma$ and its corresponding critical value $\gamma^{\mbox{\tiny{ud}}}_{\mbox{\tiny{crit}}}$. As one can observe, two clearly different regimes are found, defining two distinct flapping states of the fiber, with a well defined threshold around $\gamma=\gamma^{\mbox{\tiny{ud}}}_{\mbox{\tiny{crit}}}$. For $\gamma < \gamma^{\mbox{\tiny{ud}}}_{\mbox{\tiny{crit}}}$, i.e. the subcritical cases, all the points lay on a horizontal line characterized by an oscillation frequency equal to the turbulent one; on the contrary, for $\gamma > \gamma^{\mbox{\tiny{ud}}}_{\mbox{\tiny{crit}}}$, supercritical regime, all the points collapse on an inclined line with slope $0.5$, indicating that the oscillation frequencies grow with the bending rigidity and are larger than the turbulent frequency. These findings thus confirm the idea that the subcritical case is fully governed by turbulence, unlike the supercritical one which exhibits a structural response behavior. The net demarcation between the two regimes also allows us to quantitatively separate a hard regime of oscillation (for $\gamma > \gamma^{\mbox{\tiny{ud}}}_{\mbox{\tiny{crit}}}$) from a soft one occuring for $\gamma < \gamma^{\mbox{\tiny{ud}}}_{\mbox{\tiny{crit}}}$.

\begin{figure*}
  \centering
  \includegraphics{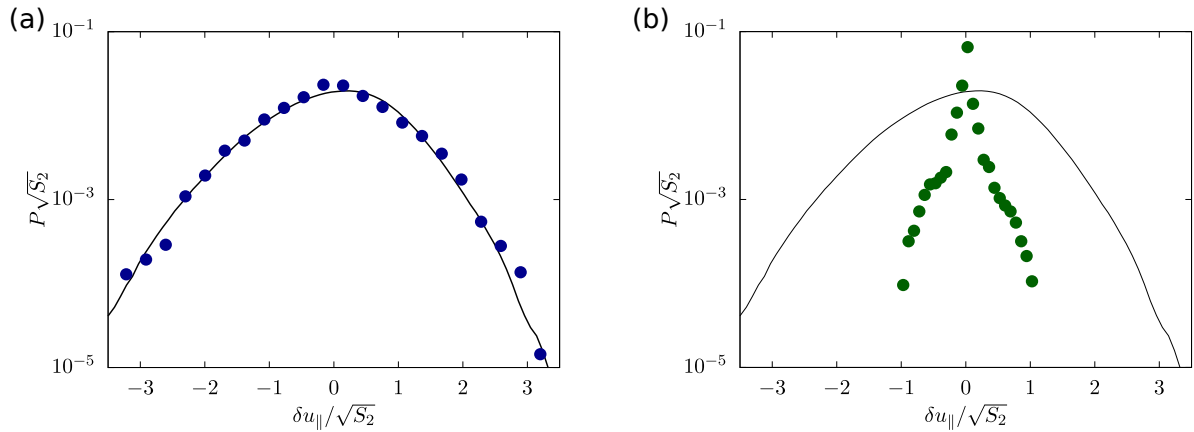}
  \caption{Probability density function (PDF) of the longitudinal velocity increments for the subcritical case $\gamma / \gamma^{\mbox{\tiny{ud}}}_{\mbox{\tiny{crit}}} = 0.3$ (left) and the supercritical case $\gamma / \gamma^{\mbox{\tiny{ud}}}_{\mbox{\tiny{crit}}} = 50$ (right) in the under-damped regime. Comparison between the Lagrangian fiber measurement (bullets) and the Eulerian one (filled curve).}
  \label{fig:pdf_ud}
\end{figure*}
A complementary analysis is performed by looking at the probability density function (PDF) of the longitudinal velocity difference $\delta u_\parallel \equiv [\mathbf{u}(\mathbf{r},t)-\mathbf{u}(\mathbf{0},t)] \cdot \hat{\mathbf{r}}$ sampled at the free ends of the fiber, compared with the same quantity computed in the Eulerian frame. Results for the two fibers considered previously are shown in Fig.~\ref{fig:pdf_ud}: the PDF of the Eulerian data shows a non-symmetric bell shape, while two very different curves are found for the two fibers. Indeed, while we notice a good agreement in the subcritical case between the Eulerian and Lagrangian data, a significative difference is found for the supercritical one. This result supports one more time the idea that the fiber dynamics in the subcritical regime is dominated by turbulence, while in the supercritical one the fiber response comes from its structural elasticity. Concerning the first case, we ascribe the small mismatch occurring between the Lagrangian and Eulerian measurements to the different number of samples (about two order of magnitudes lower for the former).

\begin{figure}
  \includegraphics[width=0.45\textwidth]{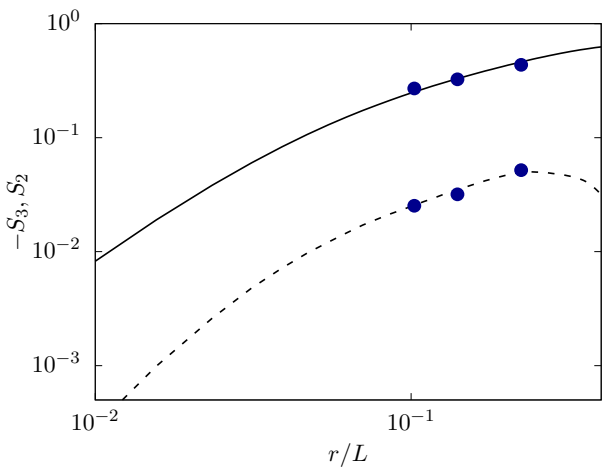} 
  \caption{Second-order (solid line) and third-order (dashed line) velocity structure functions computed in the standard Eulerian way and the same measure obtained by means of Lagrangian fiber tracking for the subcritical, under-damped case with $\gamma / \gamma^{\mbox{\tiny{ud}}}_{\mbox{\tiny{crit}}} = 0.3$ (blue bullets). Lengths and velocities are made dimensionless with the box size $L$ and with the velocity root mean square, respectively.}
  \label{fig:S23}
\end{figure}
Focusing on the subcritical situation and aiming at exploiting the capability of flexible fibers acting as a proxy of turbulent eddies, the consequent step is to use fibers of different length to obtain the flow velocity structure functions and estimate the associated scaling laws. To this end, Fig.~\ref{fig:S23} shows the second and third-order longitudinal velocity structure \sloppy{functions} $S_p(r)$ (with $p = 2, 3$), proposing a comparison between the two approaches analogous to what was presented for the PDF. Notice that the plot abscissa refers to the time-averaged value of the end-to-end distance and not to the fiber rest length, this quantity being more representative of the actual fiber length scale (see Fig.~\ref{fig:def}a). For both the reported quantities, the Lagrangian and Eulerian measurements reveal a close resemblance with differences well within the statistical error. This leads to the conclusion that, in the under-damped regime, it is possible to measure the two-point statistics of the flow (e.g., PDF, structure functions) by means of dispersed flexible fibers tracked in time, provided that  $\gamma / \gamma^{\mbox{\tiny{ud}}}_{\mbox{\tiny{crit}}} < 1$.

\subsection{Over-damped regime}
We now turn our attention to the case where $\zeta>1$ (over-damped regime), where dissipation becomes dominant. Once deformed, the fiber response is characterized by a relaxation timescale that can be estimated by balancing the elastic and viscous terms, yielding:
\begin{equation}
\tau_{\mbox{\tiny{re}}} \approx \frac{\mu c^4}{\gamma}.
\end{equation}
Note that this relaxation process has exponential behavior, without oscillations. Indeed, the fiber equation becomes first-order in time.

In this case, a balance condition can be imposed between the relaxation time and the eddy turnover one, i.e. $\tau_{\mbox{\tiny{re}}} \approx \tau (c)$, so that the critical value of the bending rigidity for this regime can be estimated as
\begin{equation}
\gamma^{\mbox{\tiny{od}}}_{\mbox{\tiny{crit}}}\sim \mu c^{10/3}\epsilon^{1/3}.
\label{gamma-crit-od}
\end{equation}

We shall discuss the expected behavior in the two limits also here. For $\gamma/\gamma^{\mbox{\tiny{od}}}_{\mbox{\tiny{crit}}}<1$, the relaxation is slower than the fluid forcing and thus the fiber is slaved to the turbulence. In the opposite case when $\gamma/\gamma^{\mbox{\tiny{od}}}_{\mbox{\tiny{crit}}}>1$, the fiber is appreciably deformed only by large strains, similarly to the under-damped regime. However, here elastic oscillations are not possible, and the dominant frequency in this case is expected to be the turbulent one. The fiber motion in the over-damped regime is therefore always expected to be slaved to turbulence, independently of $\gamma/\gamma^{\mbox{\tiny{od}}}_{\mbox{\tiny{crit}}}$.

\begin{figure}
  \includegraphics[width=0.45\textwidth]{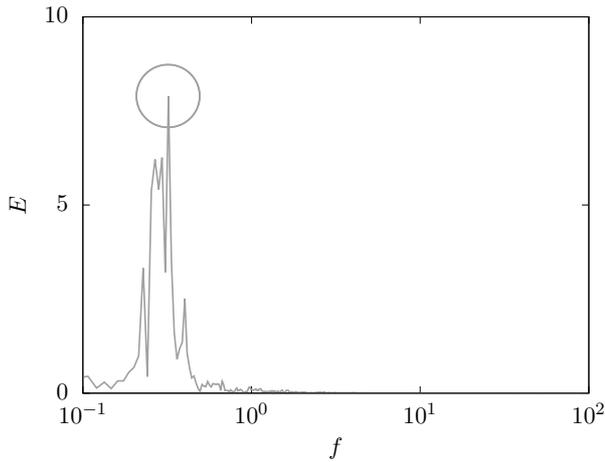}
  \caption{Temporal spectrum of the end-to-end distance for the neutrally-buoyant, over-damped case with $c/L = 0.16$ (gray line). The dominant frequency is marked with a circle.}
  \label{fig:spectraI}
\end{figure}
To prove this, we resort again to numerical experiments. When dealing with the subcritical, over-damped case, however, issues arise since the fibers are eccessively flexible, leading to numerical instability. Therefore we consider only the zero-mass case ($\rho_1  \sim 0$), corresponding to $\zeta \gg 1$. As done for the under-damped regime, the time trace of the end-to-end displacement is acquired and processed, extracting the peak of the frequency as shown in Fig.~\ref{fig:spectraI}. These data are reported in Fig.~\ref{fig:freq} (gray symbols), to show that for all computed cases the frequency ratio is approximately 1, demonstrating that the fiber flapping is locked-in to the flow.  

\begin{figure}
  \includegraphics[width=0.45\textwidth]{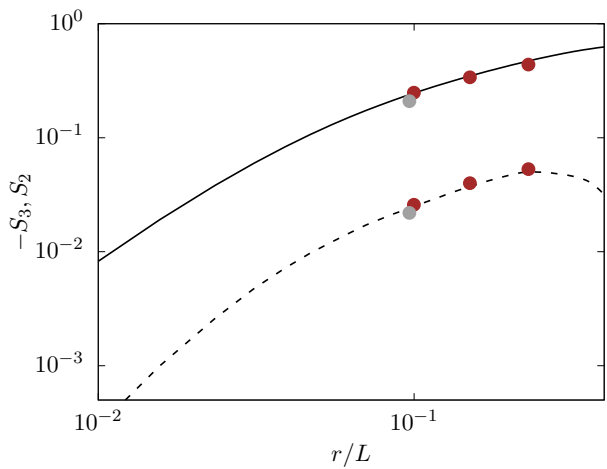}
  \caption{Second-order and third-order velocity structure functions for the neutrally-buoyant, over-damped case (gray bullets) and passive, under-damped cases (brown bullets). Lengths and velocities are made dimensionless with the box size $L$ and with the velocity root mean square, respectively.}
  \label{fig:S23p}
\end{figure}
The comparison in terms of two points turbulent statistic presented for the under-damped regime is repeated for this regime as well. First, we consider the computed velocity structure functions and the results are shown for one representative case in Fig.~\ref{fig:S23p} (along with results from the passive cases that will be introduced in Sec.~\ref{sec:passive}).
\begin{figure}
  \includegraphics[width=0.45\textwidth]{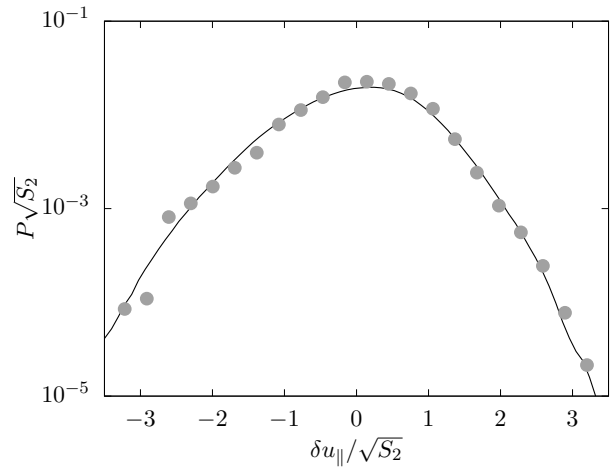}
  \caption{Probability density function (PDF) of the velocity increments for the neutrally-buoyant, over-damped case with $c/L = 0.16$. Comparison between the Lagrangian fiber measurement (bullets) and the Eulerian one (filled curve).}
  \label{fig:pdfO}
\end{figure}
Further, Fig.~\ref{fig:pdfO} depicts the resulting PDFs for the same case. Both observables measured by the Lagrangian fiber are in good agreement with the Eulerian data.

As shown for the under-damped regime, also the predictions for the over-damped case are confirmed by the results from numerical simulations.

\begin{figure}
  \includegraphics[width=0.45\textwidth]{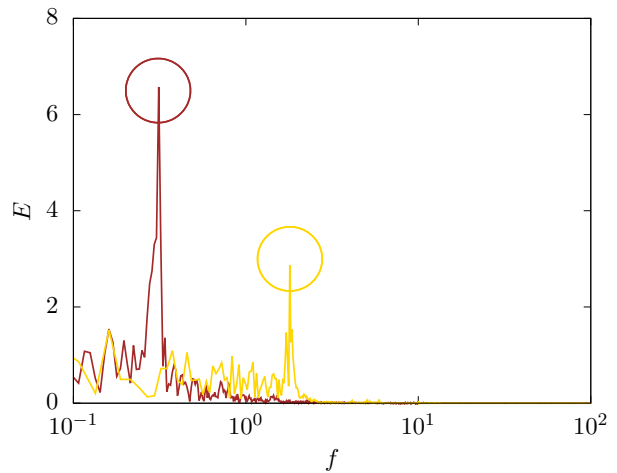}
  \caption{Temporal spectra of the end-to-end distance for subcritical case $\gamma / \gamma^{\mbox{\tiny{ud}}}_{\mbox{\tiny{crit}}} = 0.3$ (brown) and supercritical case $\gamma / \gamma^{\mbox{\tiny{ud}}}_{\mbox{\tiny{crit}}} = 50$ (yellow) in the under-damped regime, with deactivation of the fiber-flow feedback (passive cases). The dominant frequencies are marked with a circle.}
  \label{fig:spectraP}
\end{figure}

\begin{figure*}
\centering
 \includegraphics{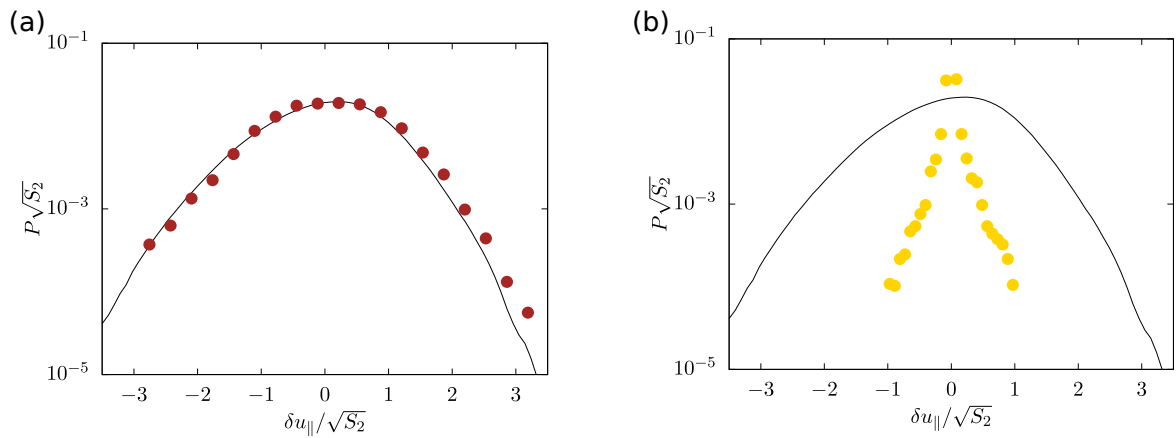}
\caption{Probability density function (PDF) of velocity increments for the passive, (a) subcritical and (b) supercritical under-damped cases with $c/L=0.16$. Comparison between the Lagrangian fiber measurement (bullets) and the Eulerian one (filled curve).}
  \label{fig:pdfP}
\end{figure*}

\section{Passive fiber}
\label{sec:passive}
In the numerical framework considered so far, the presence of the fiber modifies locally the flow. In our phenomenological model, on the other hand, we have assumed that the former has an essentially passive behavior. The question that rises is therefore: are our findings confirmed if the fiber-flow feedback in the numerical method is deactivated? To address this point, simulations are performed neglecting the feedback to the flow.

The obtained picture is the same of what found for fibers with active feedback. The spectra from which the main flapping frequency was extracted are reported in Fig.~\ref{fig:spectraP}. The data are added in Fig.~\ref{fig:freq} with brown symbols for the subcritical cases, yellow for the supercritical ones, showing that the fiber dynamics is consistent with the full model.

Focusing on the subcritical case, we examine also the velocity structure functions (Fig.~\ref{fig:S23p}, brown bullets), for which again a good agreement with the Eulerian counterpart is found. Lastly, the PDF of the longitudinal velocity difference for $c / L = 0.16$ is presented in Fig.~\ref{fig:pdfP}, again confirming the conclusions above. In light of these evidences, it appears that the action of the fiber on the flow can be neglected when modeling the dynamics of single fibers in turbulence.

\section{Suggestions for possible experiments}
\label{sec:experiments}
In this section we propose some estimations for planning laboratory experiments. To identify some possible materials for the fiber, we can refer to Ref.\ \cite{brouzet2014polymer}. For their Silicone I, the resulting bending rigidity is $\gamma = EI = 6.76 \times 10^{-7} \mathrm{Nm}^2$ and the linear density $\rho_1$ is $0.86 \mathrm{g/m}$, for their Silicone II $\gamma = EI = 1.74 \times 10^{-6} \mathrm{Nm}^2$ and $\rho_1 = 0.41 \mathrm{g/m}$, and finally, for their Nylon (III) $\gamma = EI = 2.75 \times 10^{-5} \mathrm{Nm}^2$ and $\rho_1 = 0.16 \mathrm{g/m}$. We can also refer to the silk flexible filament used in Ref. \cite{zhang2000flexible}: it is a $0.15 \mathrm{mm}$ diameter silk wire (Silk in the following) having $\gamma = EI = 10^{-10} \mathrm{Nm}^2$ and linear density $\rho_1 = 0.02 \mathrm{g/m}$. We can estimate the critical fiber length $c_{\mbox{\tiny{crit}}}$ to identify the threshold between the under-damped ($c < c_{\mbox{\tiny{crit}}}$) and the over-damped regimes ($c > c_{\mbox{\tiny{crit}}}$). Considering water as the working fluid, we have for Silicone I $c_{\mbox{\tiny{crit}}} \approx 0.15 \mathrm{m}$, for Silicone II $c_{\mbox{\tiny{crit}}} \approx 0.16 \mathrm{m}$, for Nylon III $c_{\mbox{\tiny{crit}}} \approx 0.26 \mathrm{m}$, and for Silk $c_{\mbox{\tiny{crit}}} \approx 0.0067 \mathrm{m}$. For the first 3 cases, fibers with lengths of a few centimeters (say, up to $10 \mathrm{cm}$) thus fall in the under-damped regime. For the Silk, with a centimeter-size fiber one falls in the over-damped regime, hence the fiber is expected to behave as a proxy of turbulence. Finally, for the first three fibers, we now need to estimate whether with such mechanical properties of the fiber we are in the region $\gamma < \gamma^{\mbox{\tiny{ud}}}$. To estimate whether or not this is the case, we need to know something on the turbulence field: in Ref.\ \cite{brouzet2014polymer}, their Kolmogorov scale ranges between $12$ and $91 \mathrm{\mu m}$. This implies that, in water, the largest $\epsilon$ is approximately $50 \mathrm{m^2 /s^3}$. Assuming $c = 13 \mathrm{cm}$ (that seems to be within the inertial range of the experiment) and considering the Silicone I fibers we get: $\gamma^{\mbox{\tiny{ud}}} \approx 9.95 \times 10^{-7} \mathrm{Nm^2} > 6.76 \times 10^{-7} \mathrm{Nm^2}$, yielding a ratio $\gamma/\gamma^{\mbox{\tiny{ud}}} \approx 0.68$. In 3D turbulence the under-damped regime is thus accessible.

\section{Conclusions}
\label{sec:conclusions}
This study concerns the dynamics of flexible fibers dispersed in homogeneous, isotropic turbulence. Based on simple resonance conditions between different characteristic timescales, we proposed a phenomenological model able to classify the flapping regimes experienced by the fiber. The predictions have been corroborated by fully-coupled direct numerical simulations employing an immersed boundary technique for the fluid-structure interaction.

For those regimes fully slaved to the flow, the fiber may be viewed as a Lagrangian tracker of turbulent eddies, exploitable for evaluating not only their characteristic time but also two-point statistical quantities such as, e.g., scaling exponents of velocity structure functions. We believe that this concept has potential applications in experimental methods for turbulence measurement paving the way to a new experimental strategy, a sort of ``Fiber Image Velocimetry", suitable to study turbulence two-point statistics or multipoint statistics (once a single fiber is replaced by more complex elastic structures). A substantial difference between the principle of ``Particle Image Velocimetry" and the one of the proposed ``Fiber Image Velocimetry" is that while the relative distance between a pair of tracer particles is not mantained constant in time due to the celebrated Richardson law (the relative distance between particles grows in time as $t^{3/2}$), on the other hand, this requirement is intrinsically satisfied when considering an inextensible fiber.

While in this investigation we considered only the behavior of a single, isolated fiber, the described strategy can be employed seeding the flow with several elastic objects, potentially of different lengths in order to measure eddies of different size. Increasing the concentration of the dispersed phase, however, would determine an increase of the importance of the fiber-flow feedback, so that the assumption leading to the prediction for the critical values of the bending rigidity must be modified to account for the modulation of the flow statistics caused by the fiber feed-back.

\begin{acknowledgements}
M.E.R, A.A.B and L.B. were supported by the ERC-2013-CoG-616186 TRITOS, and by the VR 2014-5001. A.M. thanks useful discussions with Alessandro Stocchino. The authors acknowledge the computer time provided by SNIC and INFN-CINECA.\end{acknowledgements}

\section*{Compliance with Ethical Standards}
The authors declare that they have no conflict of interest.

\bibliographystyle{spphys}       

\begin{thebibliography}{10}
\providecommand{\url}[1]{{#1}}
\providecommand{\urlprefix}{URL }
\expandafter\ifx\csname urlstyle\endcsname\relax
  \providecommand{\doi}[1]{DOI \discretionary{}{}{}#1}\else
  \providecommand{\doi}{DOI \discretionary{}{}{}\begingroup
  \urlstyle{rm}\Url}\fi

\bibitem{rosti_kamps_bruecker_omidyeganeh_pinelli_2017a}
M.E. Rosti, L.~Kamps, C.~Bruecker, M.~Omidyeganeh, A.~Pinelli, {M}eccanica
  \textbf{52}(8), 1811 (2017)

\bibitem{fish2006review}
F.~Fish, G.~Lauder, Annual Review of Fluid Mechanics \textbf{38}(1), 193
  (2006).
\newblock \doi{10.1146/annurev.fluid.38.050304.092201}.
\newblock
  \urlprefix\url{https://doi.org/10.1146/annurev.fluid.38.050304.092201}

\bibitem{shelley2011flapping}
M.J. Shelley, J.~Zhang, Annu. Rev. Fluid Mech. \textbf{43}, 449 (2011).
\newblock \doi{10.1146/annurev-fluid-121108-145456}.
\newblock \urlprefix\url{https://doi.org/10.1146/annurev-fluid-121108-145456}

\bibitem{lacis2014passive}
U.~L{\=a}cis, N.~Brosse, F.~Ingremeau, A.~Mazzino, F.~Lundell, H.~Kellay,
  S.~Bagheri, Nature Communications \textbf{5}, 5310 (2014)

\bibitem{peng2009}
Z.~Peng, Q.~Zhu, Physics of Fluids \textbf{21}(12), 123602 (2009).
\newblock \doi{10.1063/1.3275852}.
\newblock \urlprefix\url{https://doi.org/10.1063/1.3275852}

\bibitem{boragno2012elastically}
C.~Boragno, R.~Festa, A.~Mazzino, Applied Physics Letters \textbf{100}(25),
  253906 (2012).
\newblock \doi{10.1063/1.4729936}.
\newblock \urlprefix\url{https://doi.org/10.1063/1.4729936}

\bibitem{li2016review}
D.~Li, Y.~Wu, A.D. Ronch, J.~Xiang, Progress in Aerospace Sciences \textbf{86},
  28  (2016).
\newblock \doi{https://doi.org/10.1016/j.paerosci.2016.08.001}.
\newblock
  \urlprefix\url{http://www.sciencedirect.com/science/article/pii/S0376042116300057}

\bibitem{olivieri2017fluttering}
S.~Olivieri, G.~Boccalero, A.~Mazzino, C.~Boragno, Renewable Energy
  \textbf{105}, 530  (2017).
\newblock \doi{https://doi.org/10.1016/j.renene.2016.12.067}.
\newblock
  \urlprefix\url{http://www.sciencedirect.com/science/article/pii/S0960148116311260}

\bibitem{boccalero2017power}
G.~Boccalero, S.~Olivieri, A.~Mazzino, C.~Boragno, Smart Materials and
  Structures \textbf{26}(9), 095031 (2017).
\newblock \urlprefix\url{http://stacks.iop.org/0964-1726/26/i=9/a=095031}

\bibitem{zhang2000flexible}
J.~Zhang, S.~Childress, A.~Libchaber, M.~Shelley, Nature \textbf{408}(6814),
  835 (2000).
\newblock \doi{10.1038/35048530}.
\newblock \urlprefix\url{http://dx.doi.org/10.1038/35048530}

\bibitem{huang_shin_sung_2007a}
W.X. Huang, S.J. Shin, H.J. Sung, Journal of Computational Physics
  \textbf{226}(2), 2206  (2007).
\newblock \doi{https://doi.org/10.1016/j.jcp.2007.07.002}.
\newblock
  \urlprefix\url{http://www.sciencedirect.com/science/article/pii/S0021999107003051}

\bibitem{bagheri2012symmetry}
S.~Bagheri, A.~Mazzino, A.~Bottaro, Phys. Rev. Lett. \textbf{109}, 154502
  (2012).
\newblock \doi{10.1103/PhysRevLett.109.154502}.
\newblock
  \urlprefix\url{https://link.aps.org/doi/10.1103/PhysRevLett.109.154502}

\bibitem{voth2017anisotropic}
G.A. Voth, A.~Soldati, Annu. Rev. Fluid Mech. \textbf{49}(1), 249 (2017).
\newblock \doi{10.1146/annurev-fluid-010816-060135}.
\newblock \urlprefix\url{https://doi.org/10.1146/annurev-fluid-010816-060135}

\bibitem{sabban2017temporally}
L.~Sabban, A.~Cohen, R.~van Hout, J. Fluid Mech \textbf{814}, 42–68 (2017).
\newblock \doi{10.1017/jfm.2017.12}

\bibitem{parsa2012rotation}
S.~Parsa, E.~Calzavarini, F.~Toschi, G.A. Voth, Phys. Rev. Lett. \textbf{109},
  134501 (2012).
\newblock \doi{10.1103/PhysRevLett.109.134501}.
\newblock
  \urlprefix\url{https://link.aps.org/doi/10.1103/PhysRevLett.109.134501}

\bibitem{marchioli2010}
C.~Marchioli, M.~Fantoni, A.~Soldati, Physics of Fluids \textbf{22}(3), 033301
  (2010).
\newblock \doi{10.1063/1.3328874}.
\newblock \urlprefix\url{https://doi.org/10.1063/1.3328874}

\bibitem{ardekani2017}
M.~Niazi~Ardekani, G.~Sardina, L.~Brandt, L.~Karp-Boss, R.N. Bearon, E.A.
  Variano, Journal of Fluid Mechanics \textbf{831}, 655–674 (2017).
\newblock \doi{10.1017/jfm.2017.670}

\bibitem{parsa2014inertial}
S.~Parsa, G.A. Voth, Phys. Rev. Lett. \textbf{112}, 024501 (2014).
\newblock \doi{10.1103/PhysRevLett.112.024501}.
\newblock
  \urlprefix\url{https://link.aps.org/doi/10.1103/PhysRevLett.112.024501}

\bibitem{bordoloi2017rotational}
A.D. Bordoloi, E.~Variano, Journal of Fluid Mechanics \textbf{815}, 199–222
  (2017).
\newblock \doi{10.1017/jfm.2017.38}

\bibitem{bounoua2018tumbling}
S.~Bounoua, G.~Bouchet, G.~Verhille, Phys. Rev. Lett. \textbf{121}, 124502
  (2018).
\newblock \doi{10.1103/PhysRevLett.121.124502}.
\newblock
  \urlprefix\url{https://link.aps.org/doi/10.1103/PhysRevLett.121.124502}

\bibitem{brouzet2014polymer}
C.~Brouzet, G.~Verhille, P.~Le~Gal, Phys. Rev. Lett. \textbf{112}, 074501
  (2014).
\newblock \doi{10.1103/PhysRevLett.112.074501}.
\newblock
  \urlprefix\url{https://link.aps.org/doi/10.1103/PhysRevLett.112.074501}

\bibitem{verhille2016}
G.~Verhille, A.~Bartoli, Exp. Fluids \textbf{57}(7), 117 (2016).
\newblock \doi{10.1007/s00348-016-2201-1}.
\newblock \urlprefix\url{https://doi.org/10.1007/s00348-016-2201-1}

\bibitem{kunhappan2017}
D.~Kunhappan, B.~Harthong, B.~Chareyre, G.~Balarac, P.J.J. Dumont, Phys. Fluids
  \textbf{29}(9), 093302 (2017).
\newblock \doi{10.1063/1.5001514}.
\newblock \urlprefix\url{https://doi.org/10.1063/1.5001514}

\bibitem{rosti2018flexible}
M.E. Rosti, A.A. Banaei, L.~Brandt, A.~Mazzino, Phys. Rev. Lett. \textbf{121},
  044501 (2018).
\newblock \doi{10.1103/PhysRevLett.121.044501}.
\newblock
  \urlprefix\url{https://link.aps.org/doi/10.1103/PhysRevLett.121.044501}

\bibitem{kramel2016chiral}
S.~Kramel, G.A. Voth, S.~Tympel, F.~Toschi, Phys. Rev. Lett. \textbf{117},
  154501 (2016).
\newblock \doi{10.1103/PhysRevLett.117.154501}.
\newblock
  \urlprefix\url{https://link.aps.org/doi/10.1103/PhysRevLett.117.154501}

\bibitem{hejazi2017}
B.~{Hejazi}, M.~{Krellenstein}, G.~{Voth}, in \emph{APS Meeting Abstracts}
  (2017), p. E37.001

\bibitem{siggia_1981}
E.D. Siggia, Journal of Fluid Mechanics \textbf{107}, 375–406 (1981).
\newblock \doi{10.1017/S002211208100181X}

\bibitem{douady1991}
S.~Douady, Y.~Couder, M.E. Brachet, Phys. Rev. Lett. \textbf{67}, 983 (1991).
\newblock \doi{10.1103/PhysRevLett.67.983}.
\newblock \urlprefix\url{https://link.aps.org/doi/10.1103/PhysRevLett.67.983}

\bibitem{vergassola1997}
M.~Vergassola, A.~Mazzino, Phys. Rev. Lett. \textbf{79}, 1849 (1997).
\newblock \doi{10.1103/PhysRevLett.79.1849}.
\newblock \urlprefix\url{https://link.aps.org/doi/10.1103/PhysRevLett.79.1849}

\bibitem{falkovich2001}
G.~Falkovich, K.~Gaw\ifmmode~\mbox{\c{e}}\else \c{e}\fi{}dzki, M.~Vergassola,
  Rev. Mod. Phys. \textbf{73}, 913 (2001).
\newblock \doi{10.1103/RevModPhys.73.913}.
\newblock \urlprefix\url{https://link.aps.org/doi/10.1103/RevModPhys.73.913}

\bibitem{eswaran_pope_1988a}
V.~Eswaran, S.B. Pope, {C}omputers \& {F}luids \textbf{16}(3), 257 (1988)

\bibitem{peskin_1972a}
C.S. Peskin, {J}ournal of {C}omputational {P}hysics \textbf{10}(2), 252 (1972)

\bibitem{segel_2007a}
L.A. Segel, \emph{Mathematics applied to continuum mechanics}, vol.~52 ({SIAM},
  2007)

\bibitem{pinelli_omidyeganeh_bruecker_revell_sarkar_alinovi_2016a}
A.~Pinelli, M.~Omidyeganeh, C.~Bruecker, A.~Revell, A.~Sarkar, E.~Alinovi,
  {M}eccanica pp. 1--12 (2016)

\bibitem{rosti_brandt_2017a}
M.E. Rosti, L.~Brandt, {J}ournal of {F}luid {M}echanics \textbf{830}, 708
  (2017)

\bibitem{rosti_izbassarov_tammisola_hormozi_brandt_2018a}
M.E. Rosti, D.~Izbassarov, O.~Tammisola, S.~Hormozi, L.~Brandt, {J}ournal of
  {F}luid {M}echanics \textbf{853}, 488 (2018)

\bibitem{shahmardi_zade_ardekani_poole_lundell_rosti_brandt_2019a}
A.~Shahmardi, S.~Zade, M.N. Ardekani, R.J. Poole, F.~Lundell, M.E. Rosti,
  L.~Brandt, {J}ournal of {F}luid {M}echanics \textbf{859}, 1057 (2019)

\bibitem{Cox70}
R.G. Cox, Journal of Fluid Mechanics \textbf{44}(4), 791–810 (1970).
\newblock \doi{10.1017/S002211207000215X}

\bibitem{YS07}
Y.N. Young, M.J. Shelley, Phys. Rev. Lett. \textbf{99}, 058303 (2007).
\newblock \doi{10.1103/PhysRevLett.99.058303}.
\newblock
  \urlprefix\url{https://link.aps.org/doi/10.1103/PhysRevLett.99.058303}

\end{thebibliography}

\end{document}